# Improving Hydrogen evolution catalytic activity of 2D carbon allotrope Biphenylene with B, N, P doping: Density Functional Theory Investigations

*Mukesh Singh[a], Alok Shukla[a], Brahmananda Charkraborty[b,c,*]*

[a]Department of Physics, Indian Institute of Technology Bombay, Powai, Mumbai 400076, India

[b]High Pressure and Synchrotron Radiation Physics Division, Bhabha Atomic Research Centre, Trombay, Mumbai, India

[c]Homi Bhabha National Institute, Mumbai, India

**Abstract**

Using a first principles approach, we studied the hydrogen evolution reaction activity of newly synthesized biphenylene and B, N, P decorated biphenylene sheet. hydrogen evolution reaction activity of pristine biphenylene sheet is not encouraging, as it is similar to pristine graphene. The Gibbs free energy and overpotential of P(N) doped on biphenylene sheet are 0.022 (-0.092) eV and 22 (92) mV, respectively. The reported Gibbs free energy and overpotential of Pt are 0.9 eV and 90 mV. Hence doping of P(N) atom on top of biphenylene sheet improves hydrogen evolution reaction activity much better (near to) Pt metal. We analyzed the adsorption mechanism of dopants (B, N, P) and hydrogen with Bader charge analysis and density of states analysis. P and N-decoration on biphenylene sheet change its electronic structure so that one obtains improved hydrogen evolution reaction activity for P and N-doped biphenylene sheet. Furthermore, the stability of N, P decorated biphenylene at room temperature with ab-initio molecular dynamics and formation energy near that of biphenylene indicate experimental feasibility. We have compared all our best hydrogen evolution reaction activity results in the reaction coordinate and volcano plots of pristine, B, N, and P-doped BPh sheets. They indicate that P-doped biphenylene is a metal-free, powerful catalyst for hydrogen evolution reaction activities.

___________________________________________________________

*Corresponding author: Dr. Brahmananda ChakrabortyE-mail address: brahma@barc.gov.in
Phone: +91-2225592057

# 1. Introduction

To fulfill the required energy demands without polluting the environment, we need an alternative, independent of fossil fuels, green energy resources. Hydrogen energy with a high energy density of 142 MJ/kg [1,2], abundant in composite form, and pollution-free usage make it the most promising for future fuel resources. Unfortunately, ~ 95% of all $H_2$ produced is created through the steam reforming process in the US [3] from fossil fuels which would stop after natural gas reserves diminish, for which the estimated lifetime is ~90 years [4]. To produce sustainable and renewable and cost effective hydrogen gas, understanding the catalytic behaviors is one step essential aspect. In principle, the efficiency of hydrogen production orhydrogen-evolution reaction (HER) activity of catalysts is quantified by the Sabatier principle [5], which implies that as close the change in Gibbs free energy to zero, the lower the value of the overpotential and larger the exchange current densities. Next, we briefly discuss the present scenarios of catalysts for water splitting.

Expensive Pt groups metal (Pt, Pd, Ir, and Rh) with overpotentials around 0.1 eV are highly efficient metal catalysts [6]. Interestingly, synthesized Ru, Rh, Pd, Ir, and Pt alloys, known as high-entropy-alloy, show 7.8-9.5 times better catalytic activity than Pt [7]. Nobel metals alloys containing a small portion of them, such as Pt-Ni [8], Co-Pt [9], Pd-Ni [10], Pd-Co [11] Ru-Co [12,13], have been researched because they lead to much reduced costs. Nevertheless, even these alloys with smaller contents of Pt/Ru/Pd/Ru, etc., are not cost effective when it comes to large-scale $H_2$ production. Thus, large-scale $H_2$ production based on Pt-group metals is not sustainable both because of their scarcity and high costs.

Inspired by the superior HER catalytic performance of bio-enzymes called hydrogenases due to their active sites containing metals such Ni and Fe, the corresponding performance of transition metals Fe, Ni, and Co, their sulfides $FeS_2$, $CoS_2$, and $NiS_2$, has also been explored and found to be enhanced [14,15]. Recently, transition metal catalysts in alkaline media [16], transition metal doping on $MoS_2$ [17], transition metal carbides with N-doped carbon [18], and V-doped transition metal phosphides [19] have all been reported to speed up HER. CoN, CoP doped on carbon nanotube [20,21], $MoS_2$-qdot on MXene [22], and $SnS_2$-qdot on $MoS_2$ sheet [23] have also been shown to be effective for HER. Finally, O-terminated 2D MXenes have been found to enhance HER effectively, with the change in the Gibbs free energy comparable to the ideal case [24]. In spite of these successes, the majority of the transition-metal catalysts are not up to the standards for current industrial manufacturing processes because of their poor efficiency. Their performance could be improved by C-H functionalization, but it is limited by large amounts of catalyst loading and harsh reaction conditions, etc. Hence nanomaterials such as those based on carbon can be considered for the future design of catalysts [25].

Pristine carbon nanomaterials show negligible catalytic activity [26], similar to graphene [27,28]. However, inspired by the experimental work of Yoshikazu et al. on S/N doped on nanoporous graphene [29] with HER catalytic activity close ($\Delta G^{H*}$=0.12eV) to that of Pt, several techniques like hetero-atom doping, creation of intrinsic defects, and organics molecules absorption have been explored [30] in 2D, 1D, and 0D quantum dot carbon nanomaterials. For example, quantum dots like BN-haeckelite [31], 1D materials like graphene nanoribbon [32], g-$C_3N_4$ nanoribbons [33] and most widely investigated catalysis 2D materials [34] such N, P, S doped on graphene [27,28], single Ni-doped graphene [35], B-doped graphene [36], N/P codoped graphene [37–39], N/S codoped graphene [40,41], N/B codoped graphene [42], graphitic carbon nitride (g-$C_3N_4$) [43], $C_3N_4$ quantum dot on graphene [44], α-CN [45], $MoS_2$ [46], etc., have shown to improve HER activity.

Recently, a new allotrope of carbon, 2D Biphenylene (BPh), has been synthesized [47], and found to be mechanically robust and stable up to 4500K [48]. Its Young's modulus (212-259.7 N/m) and Poisson ratio (0.31-0.38) are much higher than Young's modulus (124 N/m) and Poisson ratio (0.25) of $MoS_2$ [49], indicating that BPh can be useful for much harsher environments as a catalyst if it has higher catalytic activity. Also, decorated BPh sheets have been reported to have high hydrogen storage capabilities [50], therefore, exploring their hydrogen production properties is of interest. Finding no prior investigations of the HER activity of BPh sheet, in this work, we have studied it for pristine BPh, B, N, and P-decorated BPh in various possible configurations. We found that Gibbs free energy $\Delta G^{H*}$ of P-decorated Biphenyle 0.022 eV surpassed that of the Pt-catalyst ($\Delta G^{H*}$=0.09). Furthermore, we have presented the DOS, PDOS, and Bader's charge transfer for the binding mechanism. Finally, all our results are compared with each other and also with that for the Pt metal using reaction coordinates and volcano plots.

## 2. Computation details:

All the calculations reported in this work were carried using first principle density-functional theory (DFT) software VASP [51,52], which implements the PAW method for pseudopotentials to obtain ground state energy for all configurations relevant for HER. We used PBE pseudopotentials in the Generalized Gradient Approximation (GGA) framework [53] and incorporate van der Waals interactions through the Grimme-D3 method [54]. Electronic configurations of valence electrons are considered $2s^22p^2$, $2s^22p^1$, $2s^22p^3$, $2s^22p^3$, and $1s^1$ for C, B, N, P, and H, respectively. Numerical parameters like energy cutoff and k-points sampling have been tested for convergence, and 500 eV and 7x7x1 Monkhrost grid [55], are found to be sufficient. Hereafter, calculations are performed with these parameters. For geometrical optimization, $10^{-6}$ eV and $10^{-2}$ eV/Å are chosen for energy and force convergence criteria. To simulate the HER activity of BPh-sheet, we have considered a 2x2x1 supercell with 20 Å vacuum. Effect of solvation are calculated with VASPsol code [56,57]. Ab initio molecular dynamics simulation [58] implemented in VASP with corrector-predictor algorithms [59] is used to verify the stability of various configurations in two stages: First, we raise the system's temperature in steps of 1fs for 5ps, where each step is assumed to be a microcanonical ensemble. Second, using the Nose-Hoover thermostat [60], we kept the system in a canonical ensemble (number of particles, volume, and temperature constant) and observed the structure for bond variation, kinetic energy, and total energy.

**Theoretical formulations:** Sabatier's principle states that interaction between catalysts and the adsorbed hydrogen must neither be too weak nor too strong, as discussed below [5]. It has been reported that hydrogen evolution reactions take place under two different reaction mechanisms: Volmer-Heyrovsky and Volmer-Tafel [61,62], which are as follows:

$$\text{Volmer reaction}: \quad H_3O^+ + M + e^- = M-H + H_2O \tag{1}$$

$$\text{Heyrovsky reaction}: \quad H^+ + e^- + M-H = H_2 + M \tag{2}$$

OR

$$\text{Tafel reaction}: \quad 2M-H = H_2 + 2M \tag{3}$$

First, the hydrogen is adsorbed on an electrode according to the Volmer mechanism, and then $H_2$ molecules form in accordance with the Heyrovsky or the Tafel mechanism. If the interaction between hydrogen and catalyst (M) is strong, H would bind strongly with M in the Volmer process ($eq^n(1)$), creating difficulties in proceeding with the Heyrovsky/Tafel reaction ($eq^n(2)$/ $eq^n(3)$), i.e., the rate of desorption would be slow. If the interaction between hydrogen and the catalyst (M) is weak, hydrogen hardly binds with the catalyst, i.e., the rate of adsorption would be slow. In Volmer, Heyrovsky, and Tafel's mechanisms, the exchange current density can be expressed in terms of change in Gibbs free energy during the reaction. From the Parson equation [63], which relates Gibbs free energy to the exchange current density, it is obvious that maximum exchange current density is obtained when the change in Gibbs free energy is zero [64].

$$\text{exchange current}: \quad i_0 = -e * k_0 \frac{1}{1+\exp(-\Delta G^{H*}/k_B T)} \tag{4}$$

Above $\Delta G^{H*}$ represents the change in Gibbs free energy during HER. In our calculation, DFT has been used to compute $\Delta G^{H*}$ using $eq^n(4)$. In equilibrium states of $eq^n(1-3)$, Gibbs free energy for HER can be computed with the following equation [28,62]

$$\text{Gibbs free energy}: \quad \Delta G^{H*} = \Delta E^{H*} + \Delta ZPE - T\Delta S \quad (5)$$

Above $\Delta E^{H*}$ is the adsorption energy of H-atom on the catalyst (BPh sheet), and $T\Delta S$ is the change of entropy of adsorbed hydrogen on BPh sheet, generally taken to be 0.24 [28,45] $\Delta ZPE$ is the change in zero point energy of adsorbed hydrogen on catalyst, which tends to be very small (0.01 to 0.04) and is therefore neglected.

$$\Delta G^{H*} = \Delta E^{H*} + 0.24 \quad (6)$$

Finally, from Gibbs free energy ($\Delta G^{H*}$), we compute the overpotential using the formula

$$\eta = \frac{|\Delta G^{H*}|}{e} \quad (7)$$

## 3. Result and discussion:

**3.1 Structural properties of BPh-sheet:** We have optimized the BPh sheet (lattice parameters a=4.52Å and b=3.76Å) using DFT and found an excellent match with the previously reported results in the literature (a=4.52Å and b=3.75Å [65]; a=4.52Å and b=3.76Å [48]). BPh is comprised of all $sp^2$ hybridized carbon, similar to graphene. However, it is more porous than graphene, with three kinds of pores: square, hexagon, and octagon. Its dynamic stability, nature of the bonding, and mechanical properties have been previously reported [48,66].

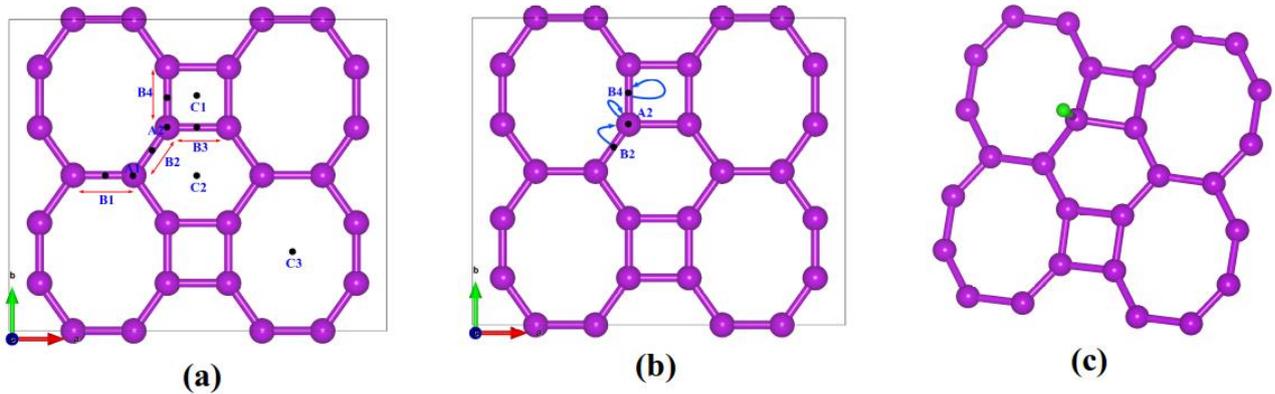

**Fig. 1:** (a) Opimized BPh sheet and possible selected positions for HER activity denoted by black dot. (b) The movement of H atom is indicated by arrow, where tail and head represent initial and final positions of H atom before and after optimization. (c) H doped on BPh sheet at A2 position. Magenta and green color represent C and H atoms.

**3.2 HER activity of BPh sheet at different sites:** We have considered all possible positions (A1, A2, B1, B2, B3, B4, C1, C2, C3) for HER activity on BPh-sheet indicated in Fig. 1(a). We initially attached a hydrogen atom to one of these sites, optimized the structure, and obtained the adsorption energy and the C-H distance of optimized structure. The final optimized positions of BPh+H are shown in Fig. 1(b), where arrows show the movement of H atom on BPh sheet after optimization. Arrow's tail points to the initial position, and the head indicates the final optimized position. Hence on doping H at A2 and B4 positions, H remains in its initial position indicated by arrows which start and end at the same position, while H atom at B2 moved to A2 position after optimization. The optimized structure of BPh+H at A2 and B2 positions are plotted in Fig. 1(c). We note that hydrogen atom doping has a negligible effect on the structures of BPh after optimization. When hydrogen is absorbed over A2, the bonds B1, B2, B3, B4 change by amount of -0.015Å, 0.098Å, 0.076Å, and 0.075Å respectively, along with a buckling of 0.56 Å presented in table S1. Similarly, doping of H over B4 position, changes bonds B1, B2, B3, and B4 by -0.009Å, 0.014Å, 0.033Å, and 0.028Å, respectively. Using eq$^n$ (5-8), Gibbs free energy ($\Delta G^{H*}$) and overpotential have been calculated and presented in table 1. From Table 1, it is obvious that A2 is the only possible site on which HER is favorable because of its small values of $\Delta G^{H*}$ (0.31 eV) and η (310 mV). In Fig. 2 (a), we have plotted the PDOS of H adsorbed at A2 position of BPh sheet, which clearly shows the hybridization of 1s orbital of H with 2p orbital of C atom of BPh sheet. Furthermore, we have presented Bader charge in table 1.

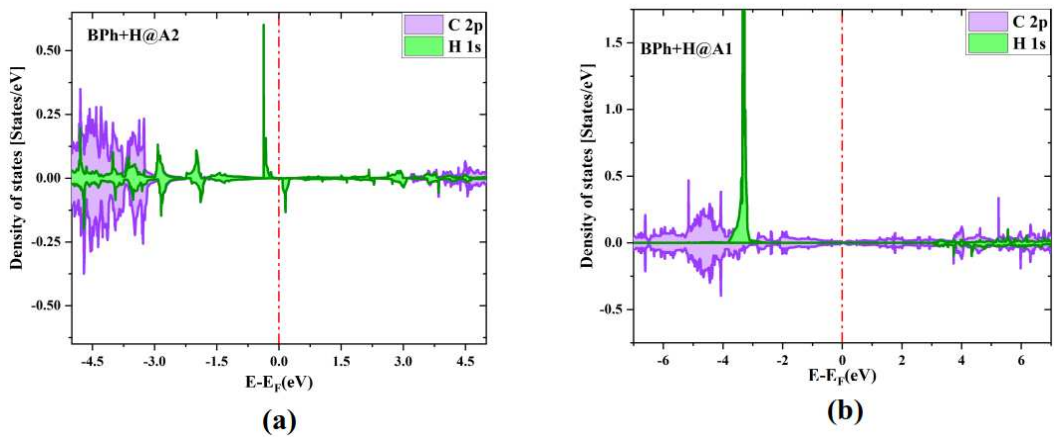

**Fig. 2:** Partial density of states (PDOS) for **(a)** H adsorbed on A2 and B2 position. **(b)** H adsorbed on A1, B1, B3, B4, C1, C2, C3 positions.

**Table 1:** Adsorption energies of H on pristine BPh sheet, change in Gibbs free energy, and overpotentials for all selected systems, along with the distances of H atoms to their nearest carbon.

| S.N. | System (@doping position) | Distance of H from nearest C (C-H) | $E_{ads}$ (eV) | $\Delta G$ (eV) | Over-potential (mV) |
|---|---|---|---|---|---|
| 1. | BPh+H(@A1) | 3.557 | 2.210 | 2.450 | 2450 |

| 2. | BPh+H(@A2) | 1.113 | 0.070 | 0.310 | 310 |
| 3. | BPh+H(@B4) | 1.351 | 1.425 | 1.665 | 1665 |

Experimentally, the performance of catalyst is quantified in terms of exchange current density and overpotentials, and better catalysts stay near the top in volcano plot and close to zero in reaction coordinate plot. Fig. 3 compares catalytic activities of theoretically ideal, Pt and pristine BPh using reaction coordinate and volcano plot. It is obvious from the plot that the best catalytic activity of BPh sheet can not perform better than Pt catalyst. Therefore, we have explored the possibility of improving BPh-sheet catalytic activity by decorating it with boron (B), nitrogen (N), and phosphorus (P) atoms.

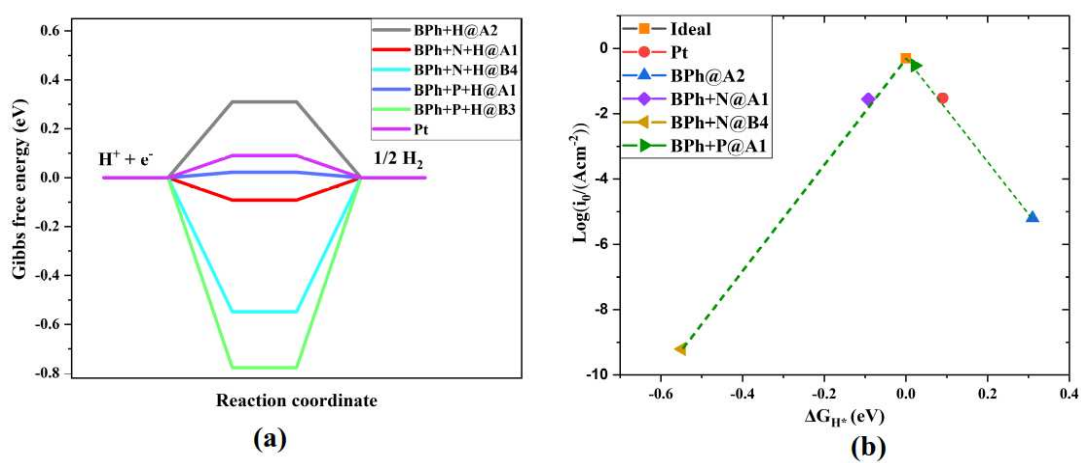

**Fig. 3:** Comparison of HER activity of our results with each other and with Pt in terms of **(a)** Gibbs free energy, **(b)** exchange current density.

**3.4 Decoration of B, N, and P on BPh-sheet, structural deformations, and Bader analysis:**

In the literature, B, N, and P atoms have been reported to improve the HER activity of carbon nanomaterial such as graphene [28]. Therefore, we decided to explore whether doping of the BPh sheet with these atoms leads to improve HER performance. The additional advantage of these atoms are that they are non-metallic therefore, the entire catalysis process remains metal free.

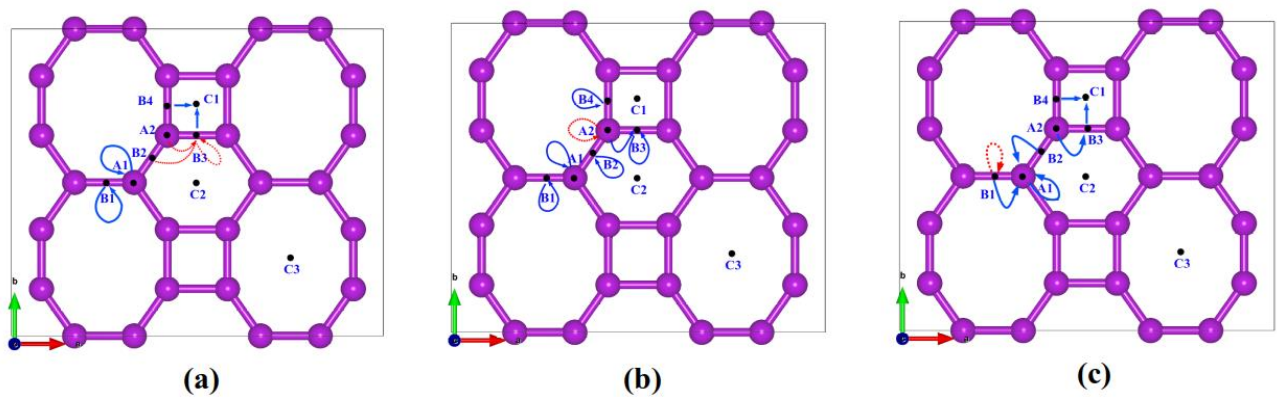

**Fig. 4:** Initial and final positions of the dopant atoms (X) before and after optimization: (a) B and H (b) N and H and (c) P and H, on BPh sheet. Solid blue curves represent initial and final structures BPh+X+H while dotted red curves show the intermediate structure BPh+X where X move to other positions after H adsorption. Where, X=B, N, P.

We decorated B, N, and P atoms at all possible sites (A1, A2, B1, B2, B3, B4, C1, C2, C3) on BPh sheet, and found that at C2 and C3 position dopants (B, N, P) did not get absorbed due to weak binding energy and large C-X distances, where X=B, N, P. For all the remaining sites, initial and final geometrically optimized structures are indicated in Fig. 4. From Fig. 4(a) indicates that B decorated on A1, B1 and B3 remain at their positions, which are represented by loop curves, starting and ending at the same sites. For A2 and B2 sites, B atom moves to B3, and for B4, B atom moves to C1. Here red dotted lines represent the local geometrically optimized structures as B would get transferred to some other site on doping H-atom (discussed in HER section). Similarly, initial and final optimized positions are indicated in Fig. 4(b) for N decorated configurations, from which it is obvious that all N atoms stay at their initial positions after geometry optimization. Finally, for P decorated BPh sheet the initial and final sites are shown in Fig. 4(c), indicating that P decorated at A1, and B1 remained at their initial positions, while P decorated at A2, B2, B3, B4 moved to B3, A1, C1, C1, respectively. All the optimized structures of B, N, and P decorated at each site are plotted in Figs. S1, S3, and S5 and their corresponding electronic structures (DOS and band structure) are plotted in Figs. S2, S4, and S6 of supporting information. The binding energy of decorated atoms and structural changes after optimization, such as bond-length of B, N, and P with respect to the nearest C atom, bond-length changes, and buckling, are presented in tables 2. Table 2 shows that changes in bond length are very small (order of $10^{-2}$ Å) except for BPh+N@B4, where C-C bond has broken after N-decoration, and new bonds are formed with N-atom (see Fig. S3(e)). However, in some cases, B, N, and P decoration on the BPh sheet led to a slight buckling. We note that the maximum buckling for B, N, and P decorated structures is 0.44 Å, 0.48 Å, and 0.52 Å, respectively, which clearly is in the ascending order of their atomic weights. Thus we can conclude if the decorating atom has large atomic weight upon absorption, it will lead to comparatively larger deformation in BPh sheet.

Bader analysis shows that on doping B on BPh sheet, B donates charges ranging from 0.31e-1.23e depending upon the position of dopant sites, leading to strong binding energies (1.5eV-2.7 eV) presented in table 2. On the contrary, for the case of decoration with nitrogen, we see from Table 2, large charge transfer from the BPh sheet to the N atom (1.5e-2.2e), except for the B4 position where C-C bond has been broken. Finally, Bader charge donated from P atoms to BPh sheet and binding energy of P atom are in the range of 0.34e-0.52e and 0.15eV-1.55eV, respectively.

**Table 2:** Distance (C-X), buckling (ΔB$_i$) and Bader charge donation from X-atom after optimization of Bph+X and binding energy of X on BPh, where X denotes boron (B), nitrongen (N), and phosphorus (P).

| Doping sites | C-X | ΔB1 | ΔB2 | ΔB3 | ΔB4 | Buckling | Bader charge donation | Binding energy (eV) |
|---|---|---|---|---|---|---|---|---|
| BPh+B | | | | | | | | |
| A1 | 1.610 | 0.070 | 0.139 | 0.054 | 0.034 | 0.180 | 1.233 | -2.220 |
| B1 | 1.871 | 0.008 | 0.015 | 0.005 | -0.018 | 0.055 | 0.315 | -1.596 |
| C1 | 1.665 | -0.008 | 0.016 | 0.021 | 0.007 | 0.094 | 0.908 | -2.746 |
| BPh+N | | | | | | | | |
| A1 | 1.533 | -0.005 | 0.062 | 0.041 | 0.064 | 0.450 | -0.625 | -1.990 |
| A2 | 1.431 | -0.014 | 0.090 | 0.081 | 0.067 | 0.482 | -0.490 | -1.506 |
| B2 | 1.427 | 0.008 | 0.086 | 0.058 | 0.040 | 0.474 | -0.620 | -1.981 |
| B3 | 1.483 | -0.007 | 0.061 | 0.066 | 0.054 | 0.435 | -0.563 | -2.257 |
| B4 | 1.331 | 0.014 | -0.008 | 0.033 | 0.577 | 0.738 | -1.244 | -3.831 |
| BPh+P | | | | | | | | |
| A1 | 1.908 | -0.005 | 0.063 | 0.048 | 0.034 | 0.523 | 0.490 | -1.260 |
| B1 | 1.960 | 0.022 | 0.033 | 0.000 | -0.030 | 0.145 | 0.341 | -0.149 |
| B3 | 2.021 | -0.014 | 0.027 | 0.027 | 0.020 | 0.187 | 0.521 | -1.547 |

### 3.5 HER activity of B-decorated BPh sheet

After finding that B atom binds strongly to the BPh sheet, we analyzed the HER activity of the BPh+B system by studying the H-doped BPh+B. We first optimized the geometry of BPh+B+H system, and using that, we computed its DOS and band structure as plotted in Figs. S7 and S8. Our calculations show that H strongly binds to the BPh+B system with binding energies ranging from 1.3 to 2.9 eV, as presented in Table 3, which indicates chemisorption of H and makes BPh+B not suitable for HER.

### 3.6 HER activity of N(P)-decorated BPh sheet:

In the next attempts, we evaluated the HER performance of N(P)-decorated BPh sheet as catalysts. Similar to B decorated BPh sheet, we optimized the geometry of BPh+N(P)+H system, and using that, we computed their DOS and band structure as shown in Figs. S9-S12. Other relevant HER parameters, such as change in the Gibbs free energy, overpotential, and the nearest distance between N(P) to C atom of BPh, and N(P) to H are presented in Table 3. From table 3, it is obvious that H adsorption on N attached to A1 leads to the HER performance quite comparable to that of Pt with the change in Gibbs free energy 0.09 eV and overpotential 90 mV. Similarly, in the case of P-decorated at A1 site of BPh, we obtained an HER performance far superior to that of Pt, with the Gibbs free energy change of 0.022 eV, and overpotential 22 mV. To consider kinetics, we have calculated the energy barrier for Volmer-Heyrovsky reactions for both BPh+N and BPh+P systems. While optimizing the reactants of Volmer reactions (eq$^n$(1)), $H_3O^+$ breaks to $H_2O$ and $H^+$ easily, and $H^+$ ion get attached to catalyst N(P) atoms. Hence, we conclude that Volmer reaction is very fast with negligible energy barriers, as in previous reports for 2D materials [67,68]. For Heyrovsky reaction (eq$^n$(2)), we optimized the reactants and products for BPh+N and BPh+P catalysts, and optimized reactants and product structures for them are plotted in Figs. 5 (b, g) and Figs. 6(b, g), respectively. Now we considered optimized reactants and products of Heyrovsky reactions as the initial and final states to create NEB images using VTST tools [69]. Then, we performed NEB calculations until the forces on all the images were optimized to be less than 0.05 eV/Å. The reaction coordinates, Figs. 5 (a) and 6 (a), for both catalysts BPh+N and BPh+P, along with optimized structures of images Figs. 5(c-f) and Figs. 6(c-f), are shown. Figs. 5(a) and 6(a) show that the kinetic energy barrier of BPh+N and BPh+P catalysis are 0.68 eV and 0.88 eV, respectively, which are comparable to 2D-$B_7P_2$ (0.64 eV) [70], Pt catalyst (0.34 eV) [71], Pt@defective graphene (0.74 eV) [72], and Pt@N-graphene(0.85 eV) [72].

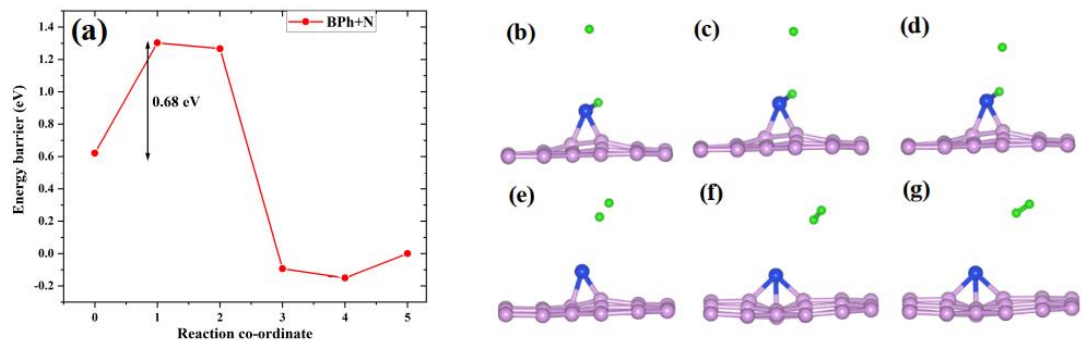

**Fig. 5:** Energy barrier of BPh+N catalyst (a), optimized initial state (b), images (c-f) and final state (g). Light magenta, blue and green spheres represent carbon, nitrogen, and hydrogen atoms.

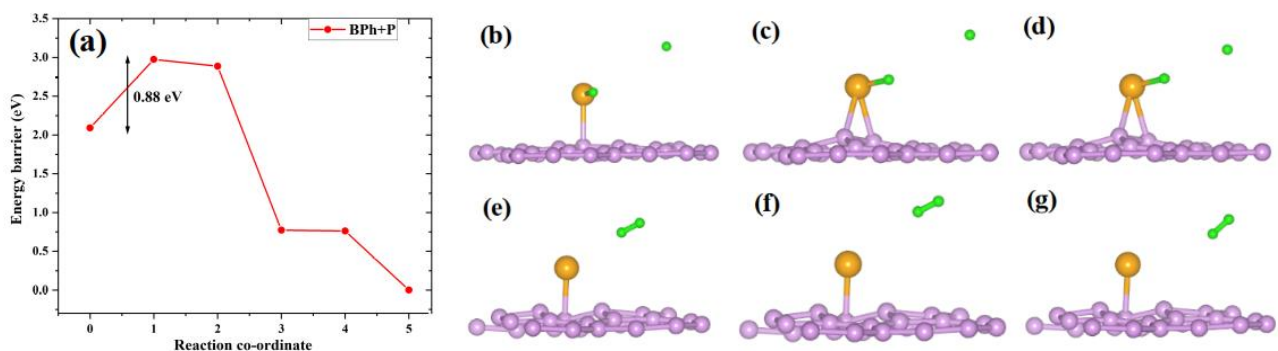

**Fig. 6:** Energy barrier of BPh+P catalyst (a), optimized initial state (b), images (c-f) and final state (g). Light magenta, dark yellow, and green spheres represent carbon, nitrogen, and hydrogen atoms

**Table 3:** Hydrogen atoms adsorption energy on X-decorated BPh sheet, change in Gibbs free energy, overpotentials (η) for all selected systems along with the distance between X and H (X-H) and distance of X atom to its nearest carbon (C-X).

| Doping sites | C-B | B-H | B.E.(eV) | $\Delta G$ (eV) | η(mV) |
|---|---|---|---|---|---|
| BPh+B+H | | | | | |
| A1 | 1.630 | 1.188 | -1.854 | -1.614 | 1,614 |
| B1 | 1.601 | 1.179 | -1.563 | -1.323 | 1323 |
| B3 | 1.661 | 1.181 | -2.831 | -2.591 | 2591 |
| C1 | 1.662 | 1.181 | -2.553 | -2.313 | 2313 |
| BPh+N+H | | | | | |
| A1 | 1.640 | 1.030 | -0.332 | -0.092 | 92 |
| A2 | 1.498 | 1.027 | -2.473 | -2.233 | 2233 |
| B2 | 1.462 | 1.028 | -1.795 | -1.555 | 1555 |
| B3 | 1.498 | 1.026 | -1.722 | -1.482 | 1482 |
| B4 | 1.361 | 1.018 | -0.789 | -0.549 | 549 |
| BPh+P+H | | | | | |
| A1 | 1.980 | 1.430 | -0.218 | 0.022 | 22 |
| B1 | 1.982 | 1.431 | -1.328 | -1.088 | 1088 |
| B3 | 1.914 | 1.439 | -1.017 | -0.777 | 777 |

**3.7 Van der Waal interaction and solvation effect:** We included the weak van der Waal interaction using Grimme-D3 correction to PBE calculated results and presented in table S3. We note that absorption energy and overpotentials for BPh+N(P) catalysts change from -0.332 (-0.218) to -0.370 (-0.250) and from 90 mV (22 mV) to 132 mV (7.5 mV), respectively. Finally, we considered our best catalyst, BPh+P, to see the solvation effect using VASPsol code. The effect of solvents on BPh+P catalyst are compared using HER activity with and without solvents in Table S4. VASPsol describes solute with a cavity surrounded by the continuum dielectric description of solvent. We have used solvents $H_2O$, HCl, and NaOH with dielectric constants of 78.4, 4.6, and 57.5, respectively. From Table S4, it is evident that the overpotentials of BPh+P in various solvents range between 17-31 mV around the overpotential of BPh+P without solvent and are better than that of Pt (90 mV) on using solvents.

## 3.8 Bader charge analysis and partial density of states of BPh+X+H systems:

To compare HER catalytic activity at different positions, we have plotted the partial density of states (2p-orbital) of N and P of the systems BPh+N+H and BPh+P+H at their stable positions. From Fig. 7, we see that PDOS of BPh+N+H@A1 has the highest peak near -0.5 eV, while for BPh+P+H@A1 the first peak occurs near -0.2 eV. For the sake of comparison, we note that the highest PDOS peak of BPh+H@A1 occurs at -0.35 eV. Further, in Table S2 we present Bader charge donated by decorated B/N/P atoms and adsorbed H atoms. Finally, we have used the Gibbs free energy and overpotential of N(P)-decorated BPh sheet to draw the reaction coordinate and volcano plots, from which a superior HER performance of these systems are obvious.

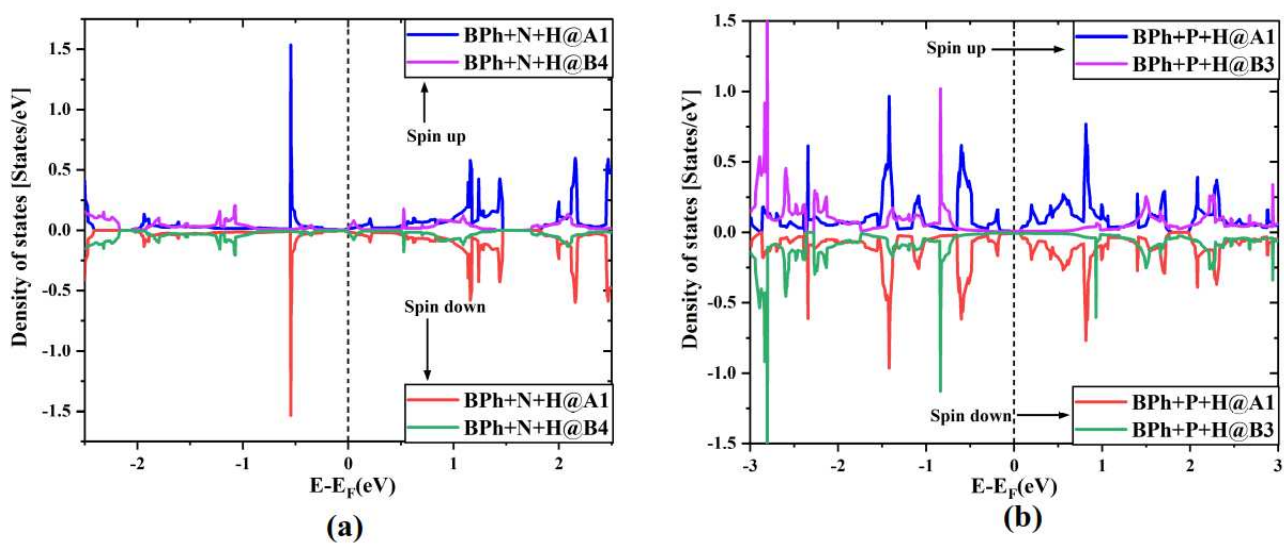

**Fig. 7:** Comparison of partial density of states of 2p-orbitals at different position of: (a) BPh+N+H, (b) BPh+P+H.

## 3.9 Feasibility of system with AIMD Simulation of N/P-decorated BPh sheet:

The cohesive energies of N, P-decorated BPh sheet are -7.17 eV and -7.14 eV, respectively, which are close to that of BPh sheet (-7.40 eV) [48], indicating the experimental feasibility of N and P-decorated systems. Using the method of incipient wetness impregnation, Pt single atoms are doped on N-doped graphene, g-$C_3N_4$, $MoS_2$ experimentally [73,74], So similar methods might be used to synthesize N(P)-decorated BPh sheet. Finally, to ensure the stability of our best catalysts, namely BPh+N@A1 and BPh+P@A1, we have performed AIMD simulations in two steps: First, we raise the system's temperature from 0K to 300K with the steps of 1fs for 5ps keeping each step in a microcanonical ensemble. Second, with the help of the Nose-Hoover thermostat, we keep the system's temperature constant at 300K and observe their structural changes (variation of bond length), kinetic energy, and total energy at every step. We have plotted the final snapshots of AIMD simulation along with their bond length variations, kinetic and total energies in Fig. 8. Figs. 8(a, b) reveal BPh+N(P) structures remain intact, while Fig.s 8(c, d) demonstrate negligible bond variations, and Figs. 8(e, f) exhibit negligible kinetic and total energy fluctuations. Consequently, the data from Fig. 8 indicate that throughout the finite temperature AIMD simulations, systems BPh+N(P) are thermodynamically stable.

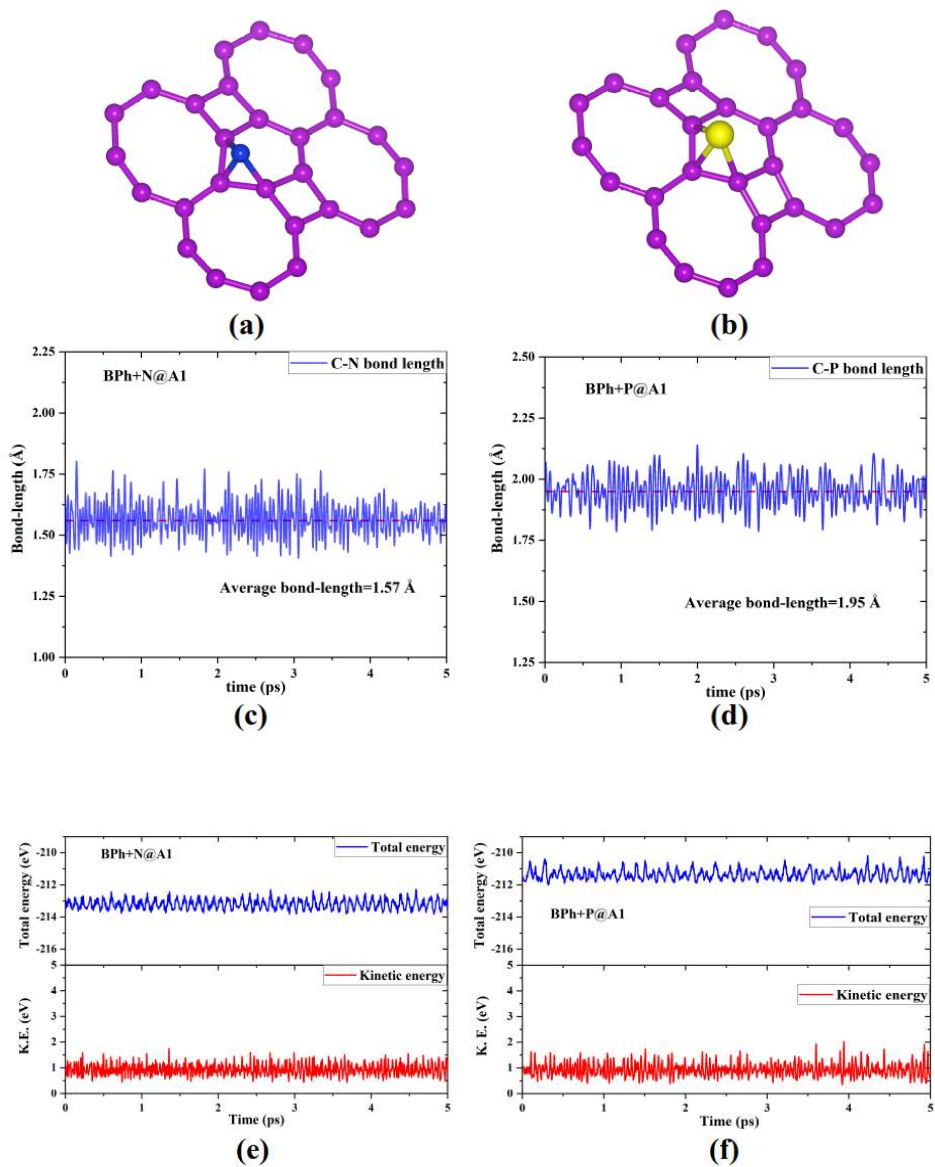

*Fig. 8:* AIMD snapshot of *(a)* BPh+N and *(b)* BPh+P after 5ps; Variation of *(c)* bond length C-N of BPh+N *(d)* bond length C-P of BPh+P *(e)* Kinetic and total energy BPh+N *(f)* Kinetic energy and total energy of Bph+P during AIMD of 5ps with steps of 1fs.

**4. Conclusion:** In summary, we have performed DFT calculations of pristine and B, N, and P decorated BPh sheets. For the purpose, we computed the change in Gibbs free energy during the HER, which relates to the overpotential and exchange current density. We have plotted Gibbs free energy and exchange current density in the reaction coordinate and volcano plot to compare the

HER activity of pristine BPh, B, N, and P decorated BPh, and Pt, with theoretically ideal catalyst corresponding to $\Delta G^{H*}$=0. Using small overpotential as a criterion for efficient HER, biphenylene is found to be ineffective similar to previous carbon nanomaterials such as graphene. However, P decoration on BPh sheet increases its HER activity tremendously, with an overpotential of 22 mV, surpassing even Pt noble metal, which has a 90 mV overpotential. Additionally, including the solvation effect on P decorated BPh shows HER activity with overpotential (17-31mV). Furthermore, partial density of states and Bader charge analysis of our analyzed system are presented. Finally, the stabilities of our best results have been checked with AIMD simulation. We hope that metal-free P and N-decorated BPh catalysts with better HER catalytic activity would inspire the experimentalist to explore its practical feasibility.

**Acknowledgment**

MS sincerely wants to thank CSIR for supporting his Ph.D. research projects. MS also appreciates the Spacetime High Performance Computing team and its facility at IIT Bombay. BC acknowledge Dr. T. Sakuntala and Dr. Nandini Garg for their support. BC also thanks Dr. S.M. Yusuf and Dr. A. K. Mohanty for their encouragement.

# Improving Hydrogen evolution catalytic activity of 2D carbon allotrope Biphenylene with B, N, P doping: Density Functional Theory Investigations

*Mukesh Singh[a], Alok Shukla[a], Brahmananda Charkraborty[b,c,*]*

[a]Department of Physics, Indian Institute of Technology Bombay, Powai, Mumbai 400076, India

[b]High Pressure and Synchrotron Radiation Physics Division, Bhabha Atomic Research Centre, Trombay, Mumbai, India

[c]Homi Bhabha National Institute, Mumbai, India

**Geometry optimization and electronic strcuture of BPh+B:**

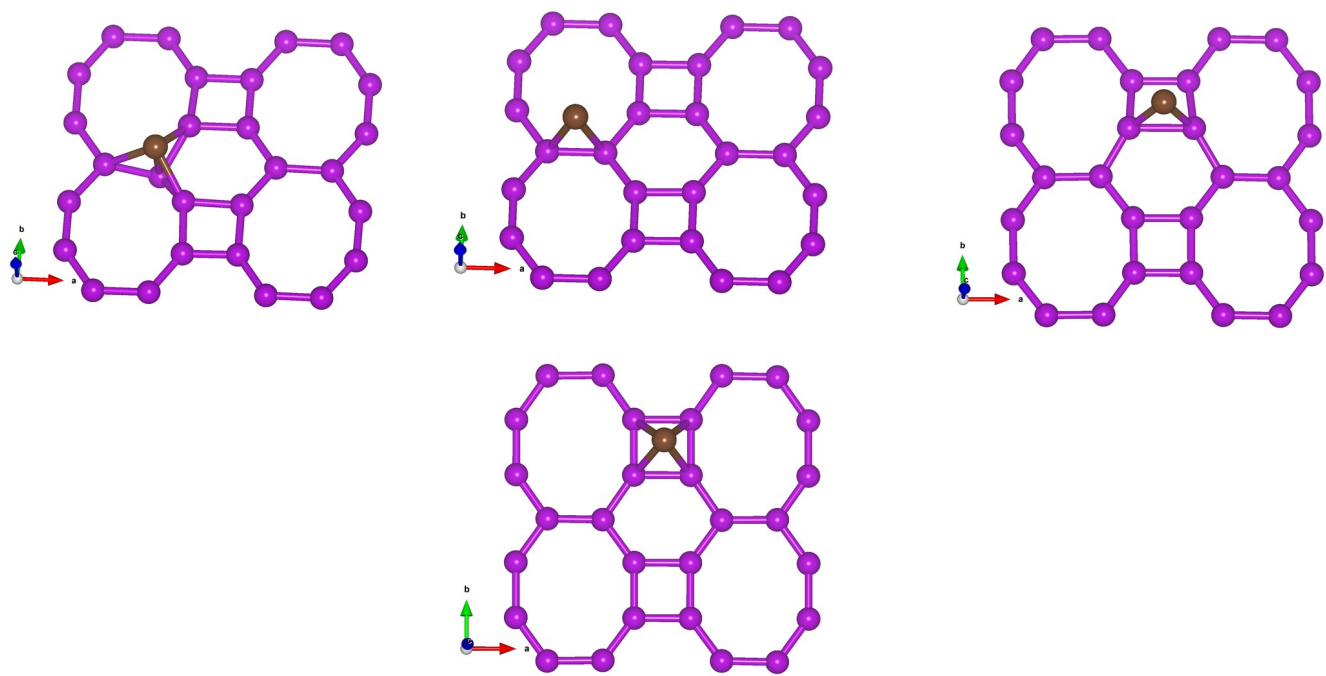

**Fig. S1:** Optimized structures of B decorated on BPh sheet at (a) A1, (b) B1, (c) B3, and (d) C1 positions.

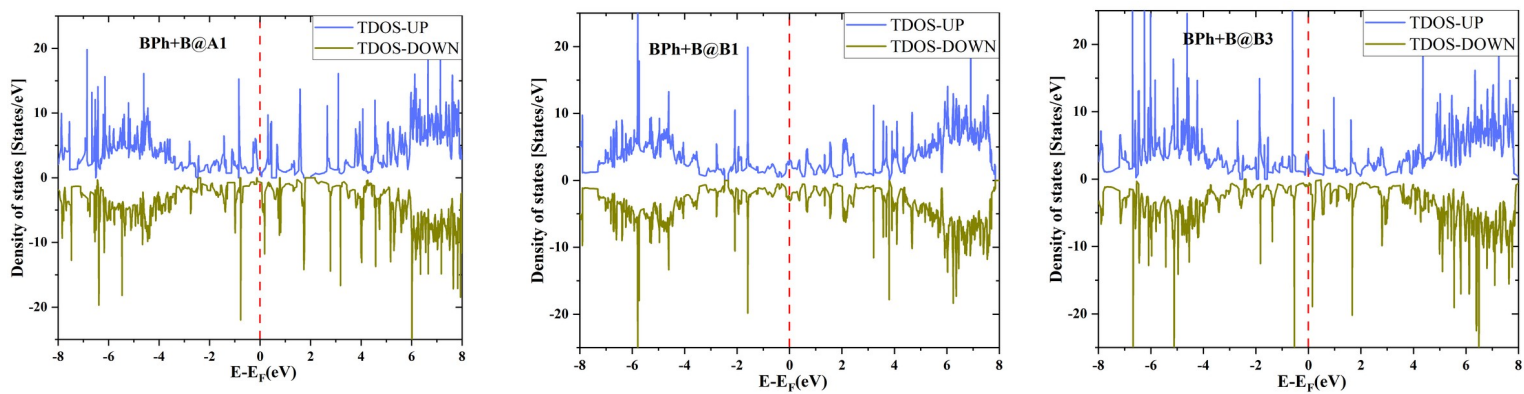

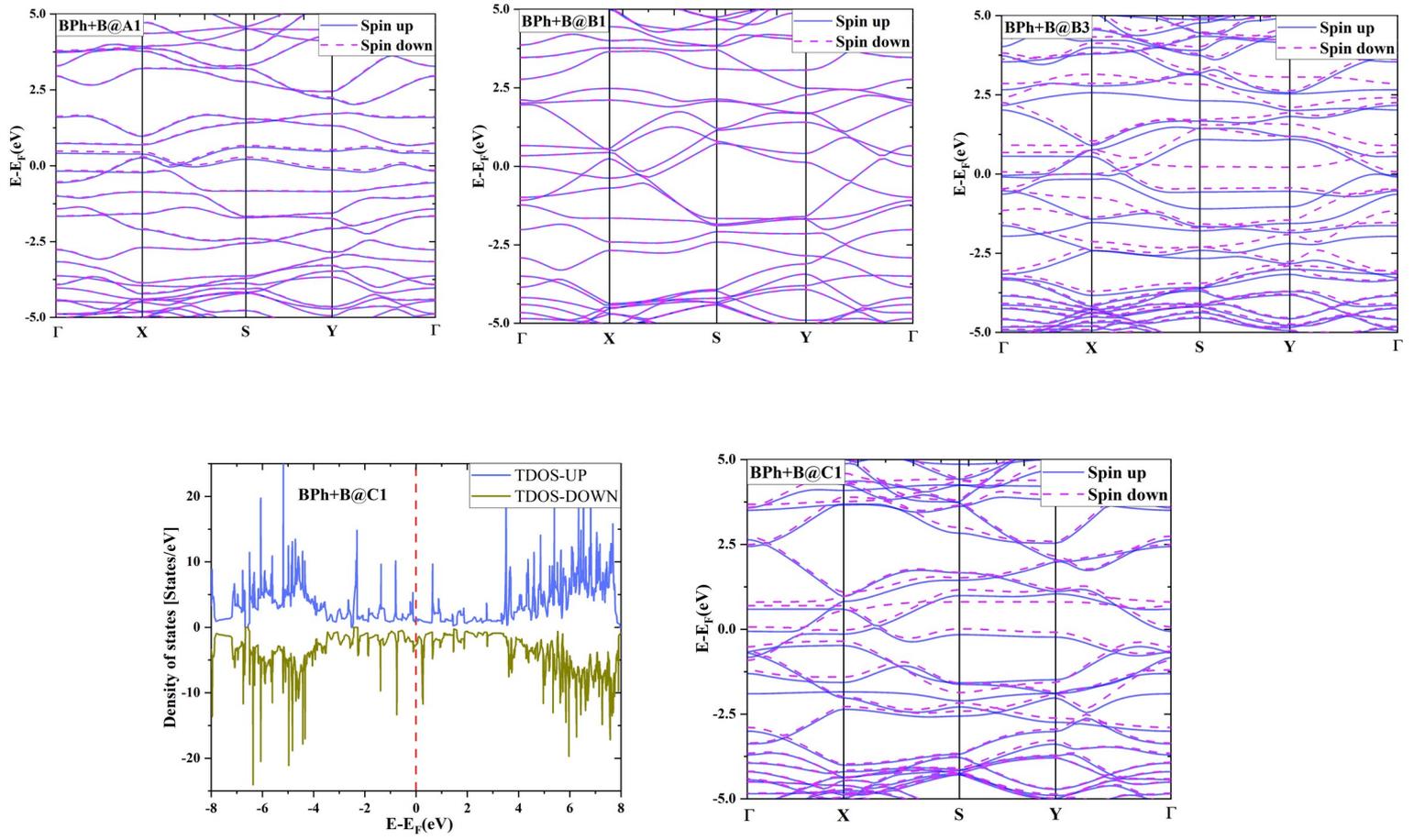

**Fig. S2:** Electronic structure of B decorated on BPh sheet for all the geometrically stable configurations. Top and left panels contain the total density of states while bottom and right panels contain the band structure.

**Geometry optimization and electronic strcuture of BPh+N:**

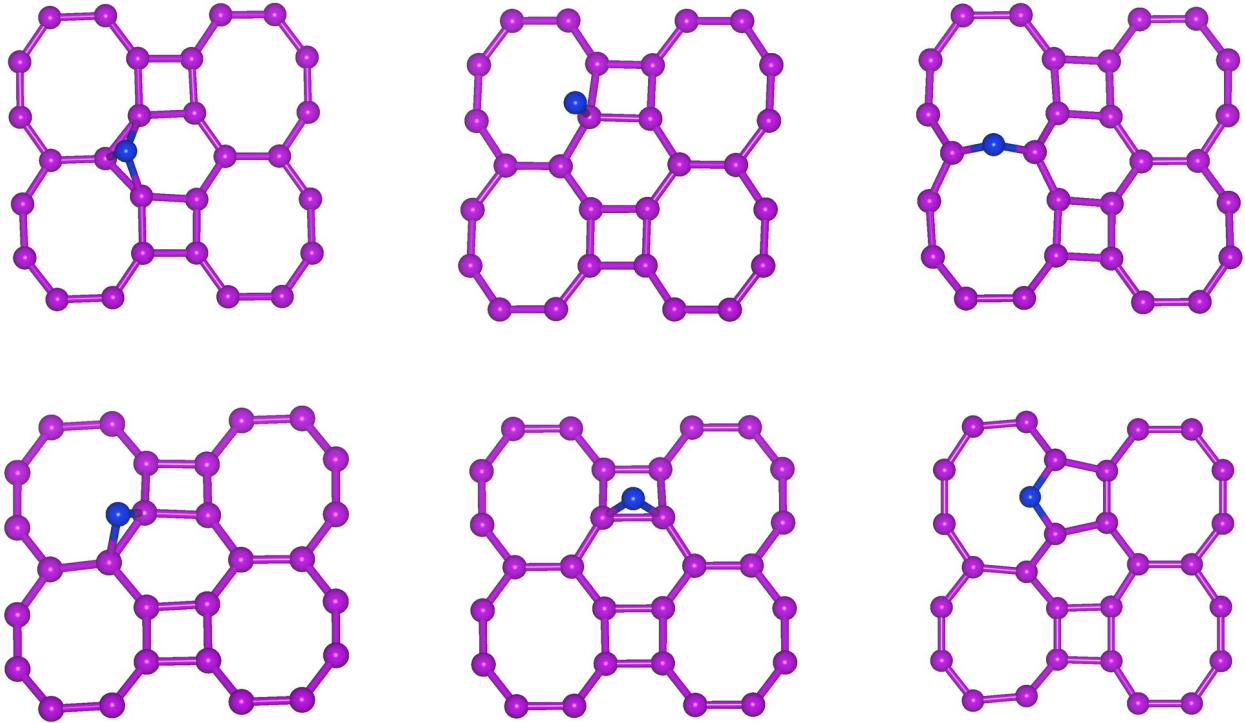

**Fig. S3:** Optimized structures of N decorated on BPh sheet at (a) A1, (b) A2, (c) B1, (d) B2, (e) B3, (f) B4 positions.

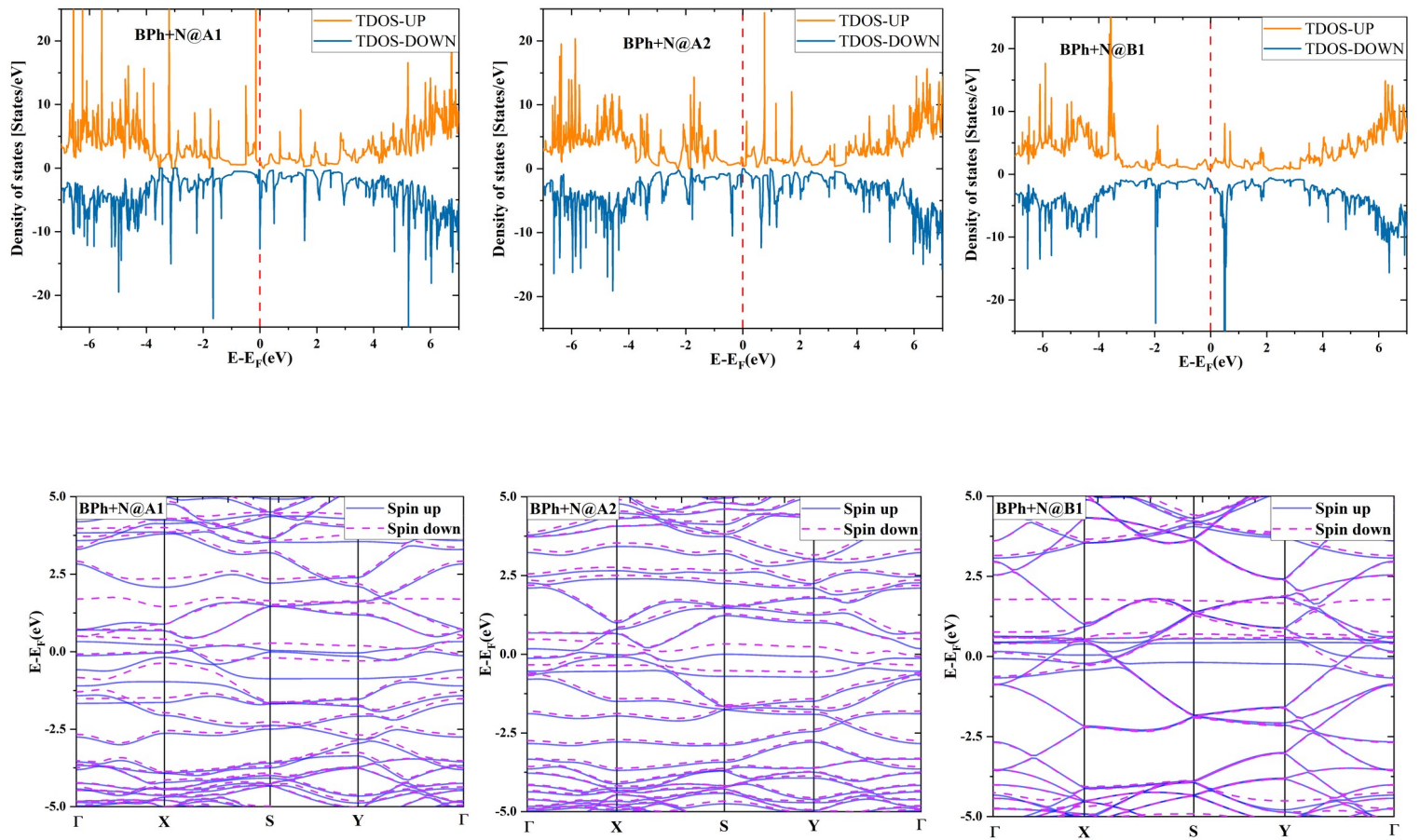

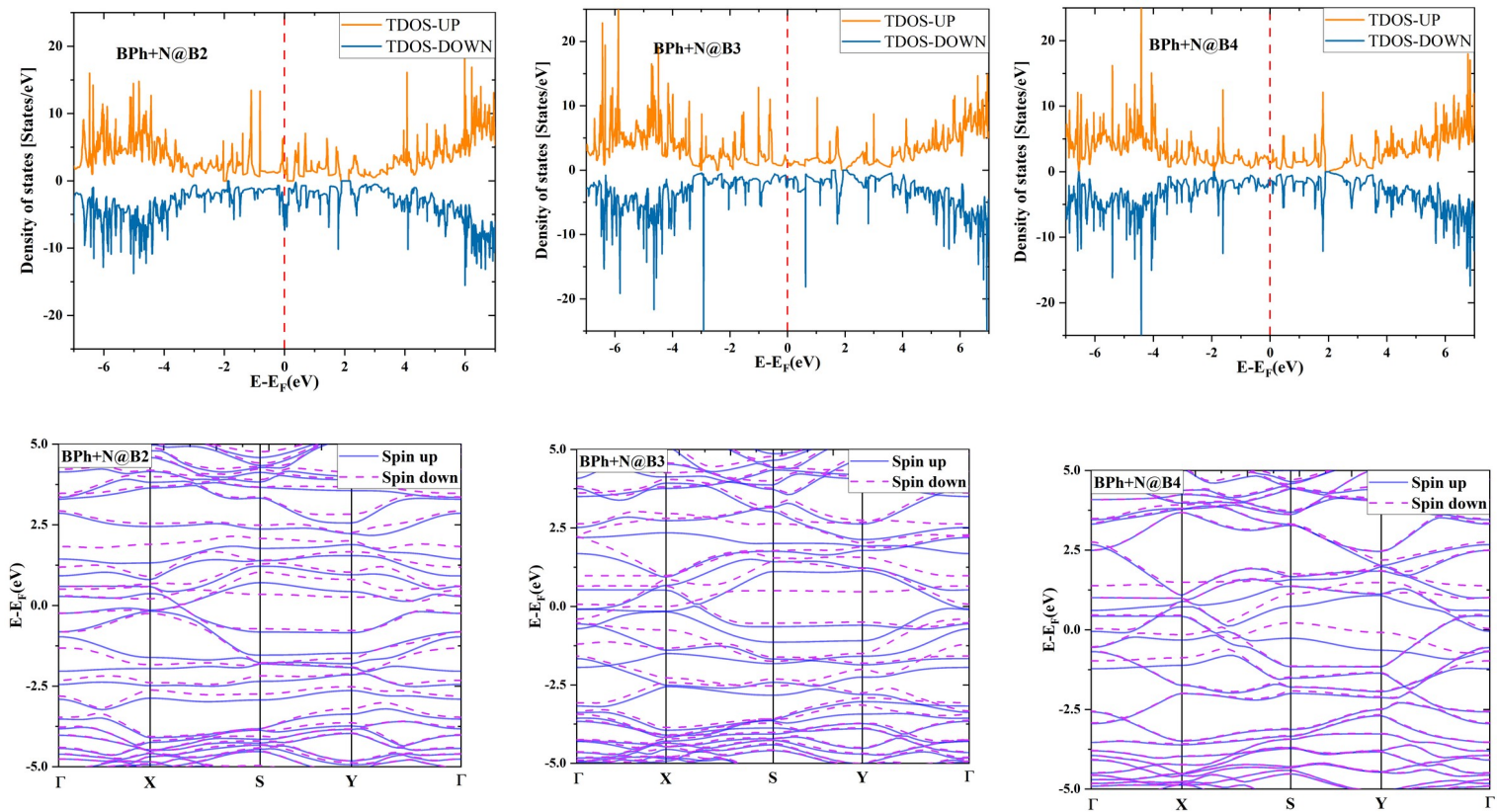

**Fig. S4:** Electronic structure of N decorated on BPh sheet for all the geometrically stable configurations. Top and bottom panels contain the total density of states and the band structure, respectively.

**Geometry optimization and electronic strcuture of BPh+P:**

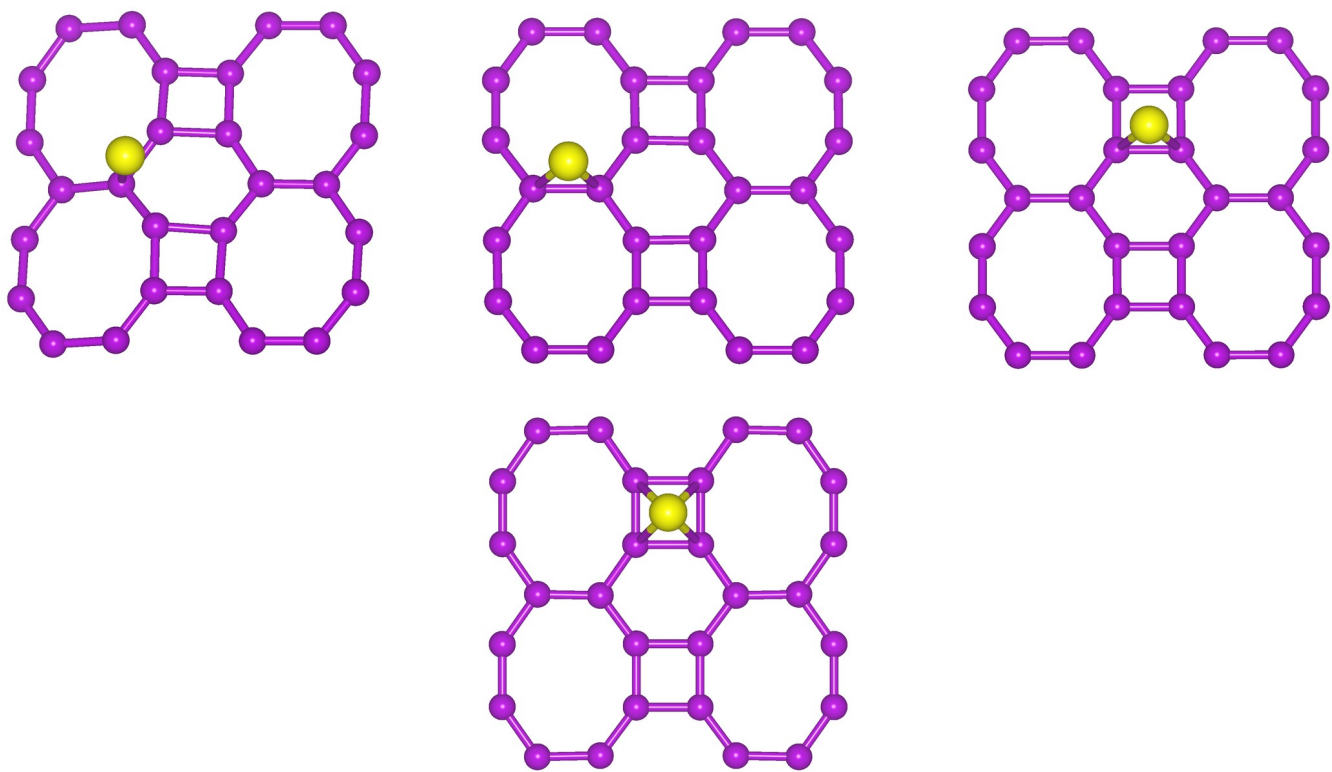

**Fig. S5:** Optimized structures of P decorated on BPh sheet at (a) A1, (b) B1, (c) B3, and (d) C1 positions.

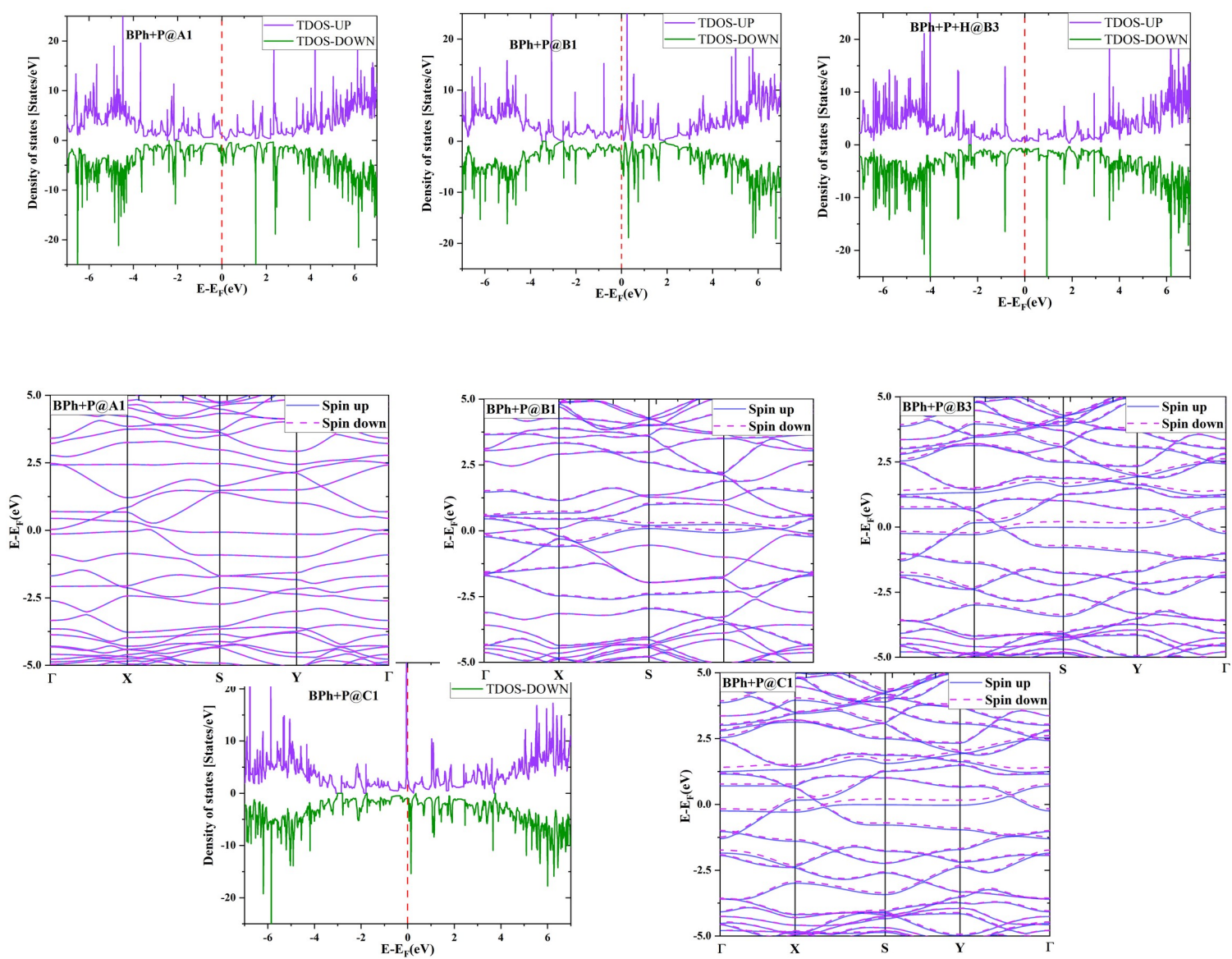

**Fig. S6:** Electronic structure of P decorated on BPh sheet for all the geometrically stable positions. Top and left panels contain the total density of states while bottom and right panels contains the band structure.

**Geometry optimization and electronic strcuture of BPh+B+H:**

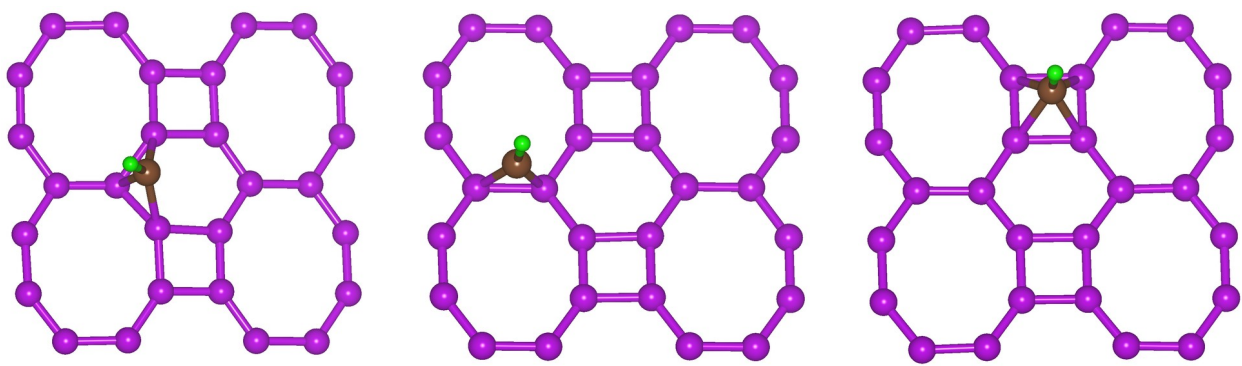

**Fig. S7:** Optimized structure of BPh+B+H sheet at (a) A1, (b) B1, and (d) C1 positions.

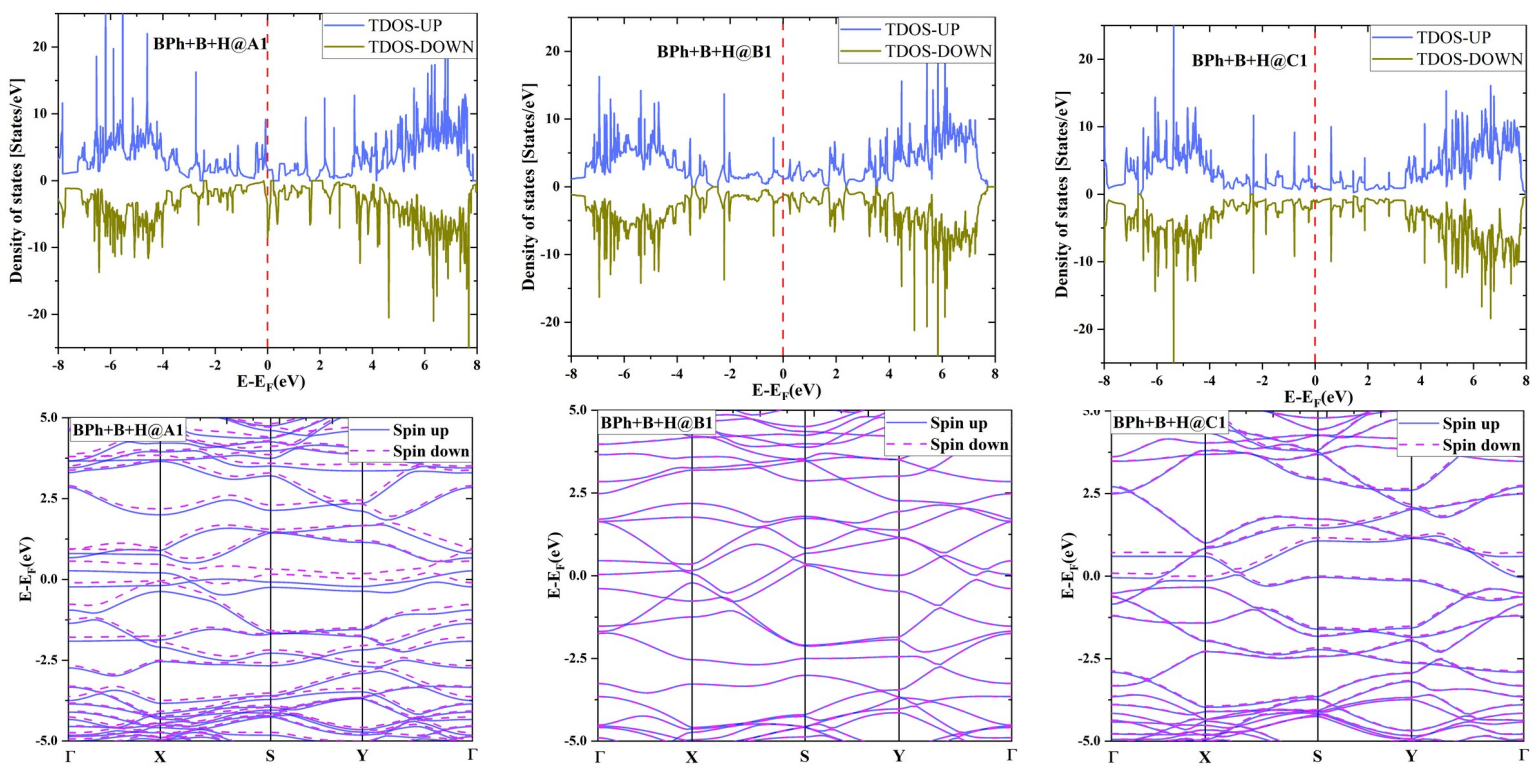

**Fig. S8:** Electronic structure of BPh+B+H sheet for all the geometrically stable configurations. Top and right panels contain the total density of states and the band structures, respectively.

**Geometry optimization and electronic strcuture of BPh+N+H:**

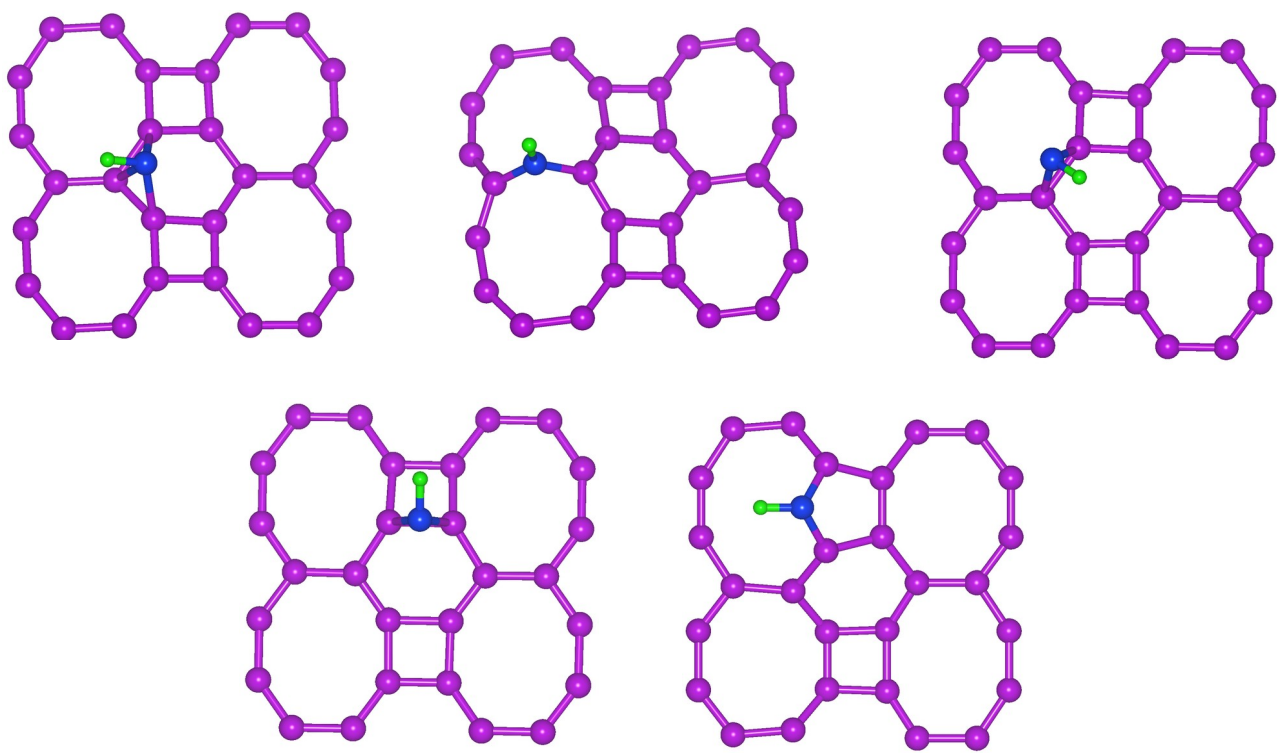

**Fig. S9:** Optimized structures of BPh+N+H sheet at (a) A1, (b) B1, (c) B2, (d) B3, and (e) B4 positions.

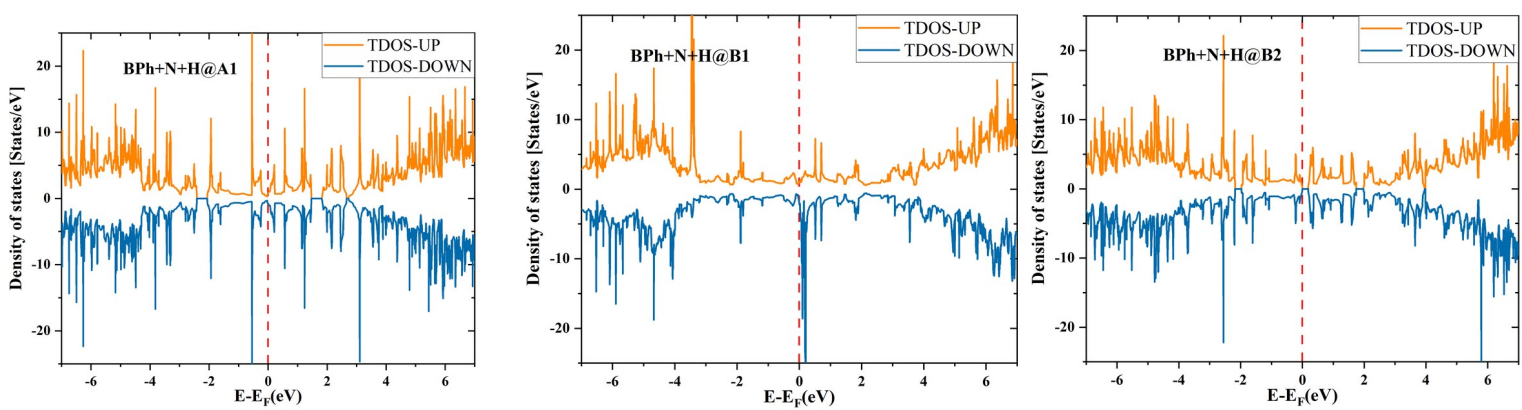

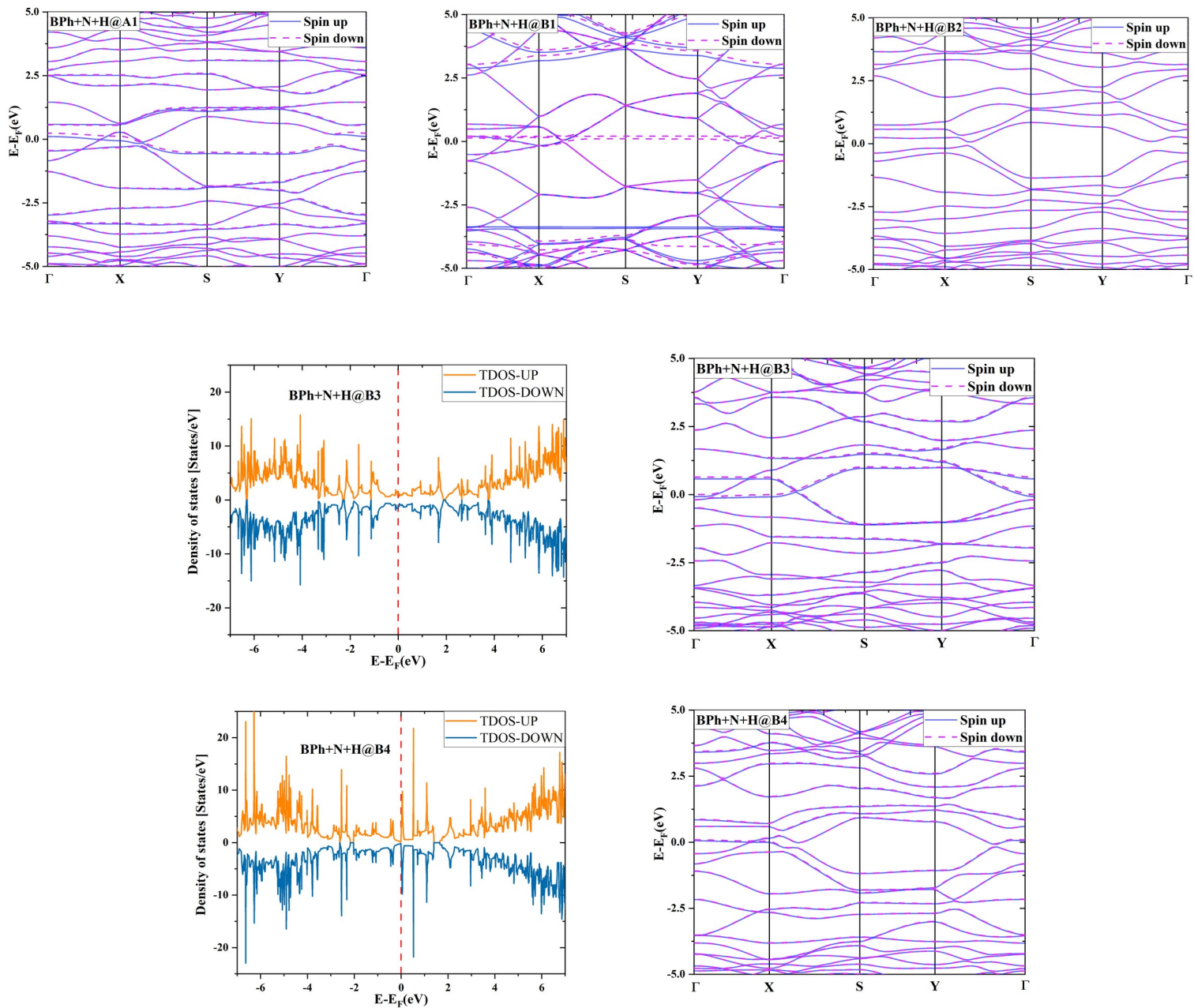

**Fig. S10:** Electronic structure of BPh+N+H for all the geometrically stable configurations. Top and left panels contain the total density of states while bottom and right panels contain the band structure.

**Geometry optimization and electronic strcuture of BPh+P+H:**

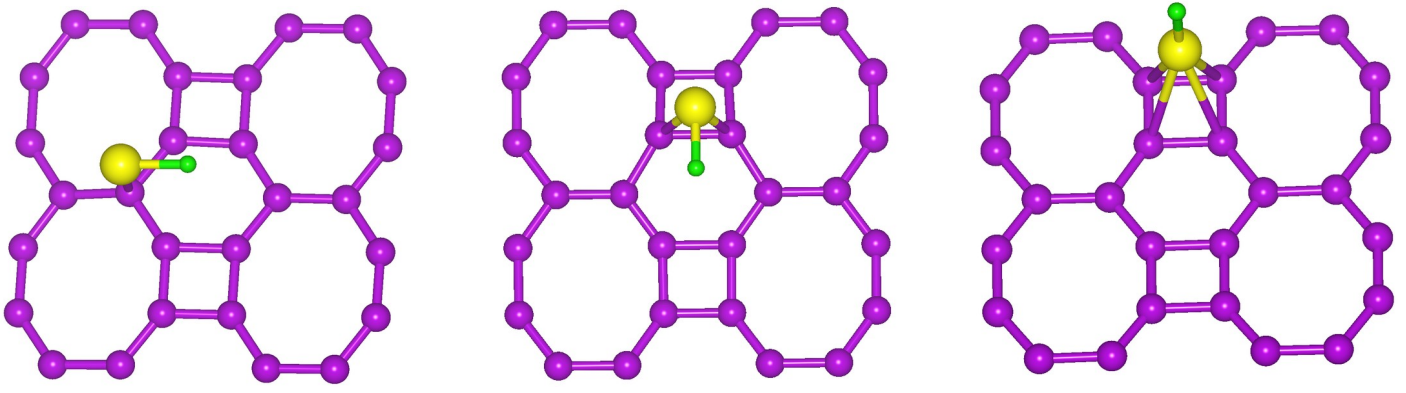

**Fig. S11:** Optimized structures of BPh+P+H sheet at (a) A1, (b) B1, and (d) C1 positions.

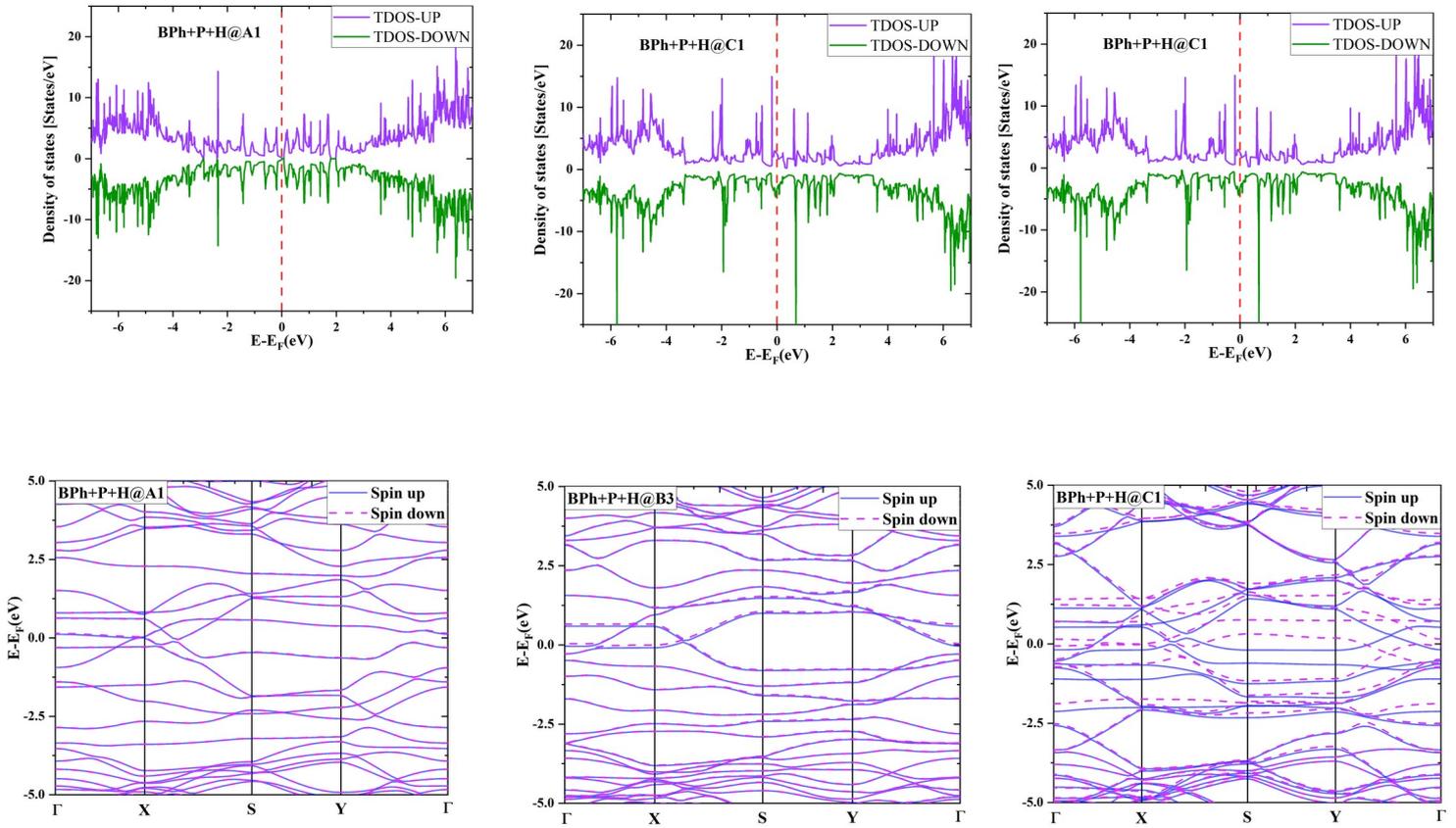

**Fig. S12:** Electronic structure of BPh+P+H for all the geometrically stable configurations. Top and right panels contain the total density of states and the band structure.

**Table S1:** Change in bond-length, buckling, and Bader charge donation from H atom after optimization of BPh+H.

| Doping sites | ΔB1 | ΔB2 | ΔB3 | ΔB4 | Buckling | Bader charge donation |
|---|---|---|---|---|---|---|
| A1 | 0.000 | 0.001 | -0.002 | 0.000 | 0.002 | 0.002 |
| A2 | -0.015 | 0.098 | 0.076 | 0.075 | 0.563 | 0.042 |
| B4 | -0.009 | 0.014 | 0.033 | 0.028 | 0.103 | 0.127 |

**Table 2:** Change in bond length, buckling and Bader charge dontated from X and H atoms, after optimization of BPh+X+H where X denotes B, N, and P atoms.

| Doping sites | ΔB1 | ΔB2 | ΔB3 | ΔB4 | Buckling | Bader charge ||
|---|---|---|---|---|---|---|---|
| | | | | | | X-donated | H-donated |
| BPh+B+H ||||||||
| A1 | -0.028 | 0.083 | 0.060 | 0.031 | 0.402 | 0.396 | -0.540 |
| B1 | 0.033 | 0.026 | 0.013 | -0.032 | 0.123 | 1.242 | -0.542 |
| B3 | -0.012 | 0.027 | 0.034 | 0.000 | 0.044 | 0.582 | -0.477 |
| C1 | -0.012 | 0.027 | 0.035 | 0.000 | 0.054 | 0.690 | -0.471 |
| BPh+N+H ||||||||
| A1 | -0.018 | 0.063 | 0.041 | 0.042 | 0.393 | -0.086 | 0.451 |
| A2 | -0.006 | 0.077 | 0.065 | 0.067 | 0.537 | -0.370 | 0.405 |
| B2 | 0.022 | 0.090 | 0.048 | 0.043 | 0.514 | -0.258 | 0.409 |
| B3 | -0.006 | 0.077 | 0.065 | 0.067 | 0.541 | -0.295 | 0.407 |
| B4 | 0.019 | -0.022 | 0.001 | 0.664 | 0.777 | -0.063 | 0.493 |
| BPh+P+H ||||||||
| A1 | 0.023 | 0.038 | 0.004 | -0.011 | 0.286 | 0.060 | -0.361 |
| B1 | 0.023 | 0.038 | 0.005 | -0.011 | 0.289 | 0.243 | -0.346 |
| B3 | -0.003 | 0.071 | 0.076 | 0.060 | 0.527 | 0.372 | -0.406 |
| C1 | -0.003 | -0.004 | 0.051 | -0.020 | 0.477 | 0.313 | -0.412 |

**Table S3:** Comparison of overpotentials (η) with and without solvents for BPh+P catalyst:

| S. N. | Solvents | Overpotential (η)(mV) |
|---|---|---|
| 1. | Vacuum | 22 |
| 2. | $H_2O$ | 17 |
| 3. | HCl | 31 |
| 4. | NaOH | 20 |

**Table S4:** Using PBE+Grimme-D3, hydrogen atoms adsorption energy on X-decorated BPh sheet, change in Gibbs free energy, overpotentials (η) for all selected systems along with the distance between X and H (X-H) and distance of X atom to its nearest carbon (C-X).

| Doping sites | C-B | B-H | B.E.(eV) | ΔG (eV) | η (mV) |
|---|---|---|---|---|---|

|   |       |       | BPh+B+H |        |          |
|---|-------|-------|---------|--------|----------|
| A1 | 1.679 | 1.189 | -1.900 | -1.664 | 1663.950 |
| B1 | 1.601 | 1.178 | -1.578 | -1.338 | 1337.510 |
| B3 | 1.662 | 1.181 | -2.593 | -2.353 | 2352.680 |
| C1 | 1.661 | 1.181 | -2.592 | -2.352 | 2351.940 |
|   |       |       | BPh+N+H |        |          |
| A1 | 1.641 | 1.031 | -0.370 | -0.132 | 132.030  |
| A2 | 1.498 | 1.027 | -2.491 | -2.251 | 2250.900 |
| B2 | 1.462 | 1.028 | -1.827 | -1.587 | 1587.100 |
| B3 | 1.498 | 1.027 | -1.753 | -1.513 | 1512.940 |
| B4 | 1.362 | 1.018 | -0.831 | -0.591 | 591.480  |
|   |       |       | BPh+P+H |        |          |
| A1 | 1.983 | 1.431 | -0.250 | -0.008 | 7.590    |
| B1 | 1.983 | 1.431 | -1.374 | -1.134 | 1134.110 |
| B3 | 1.914 | 1.439 | -1.056 | -0.816 | 815.900  |
| C1 | 1.934 | 1.437 | 0.468  | 0.708  | 707.900  |